# Early Risk Prediction of Chronic Myeloid Leukemia with Protein Sequences using Machine Learning based Meta-Ensemble

Madiha Hameed1,2, Muhammad Bilal1, Tuba Majid 3,∗, Abdul Majid 1,4 and Asifullah Khan 1,4

1 Pattern Recognition Lab (PRLab), Department of Computer and Information Sciences, Pakistan Institute of
Engineering and Applied Sciences, Islamabad,
2 Department of Information Technology Services, Hafiz Hayat Campus, University of Gujrat, Punjab
3 Experimental Continuum Mechanics (ECM) Research Group, Department of Mechanical and Process
Engineering, ETH Zürich, Zürich
4 PIEAS Artificial Intelligence Center, Pakistan Institute of Engineering and Applied Sciences, Nilore, Islamabad  * Correspondence: tmajid@ethz.ch;

**Abstract**: Leukemia, the cancer of blood cells, originates in the blood-forming cells of the bone
marrow. In Chronic Myeloid Leukemia (CML) conditions, the cells partially become mature that
look like normal white blood cells but do not resist infection effectively. Early detection of CML is important for effective treatment, but there is a lack of routine screening tests. Regular check-ups
and monitoring of symptoms are the best way to detect CML in the early stages. In the study, we
developed a multi-layer-perception-based meta-ensemble system using protein amino acid sequences for early risk prediction of CML. The deleterious mutation analysis of protein sequences provides 7discriminant information in amino acid sequences causing CML. The protein sequences are expressed
into molecular descriptors using the values of hydrophobicity and hydrophilicity of the amino acids. 9
These descriptors are transformed in various statistical and correlation-based feature spaces. These 10
features information is given to several diverse types of base learners. The preliminary predictions of 11
base-learners are employed to develop Multi-Layered Perceptron (MLP) based meta-ensemble. The 12
proposed learning approach effectively utilizes the discriminant information to classify CML/non- 13
CML protein sequences. The proposed prediction system has given improved results and it can be 14
employed as a potential biomarker for early diagnosis of CML.

## 1. Introduction

Chronic Myeloid Leukemia (CML) originates in the blood-forming cells of the bone 19
marrow that spreads slowing throughout the body when the DNA of the bone marrow 20
cell is damaged. The high proliferation rate of CML cells, reduced apoptosis, inhibition 21

of differentiation, and high chemotherapeutic resistance. These are the key functional alterations linked to the progression of the disease. These alterations enable CML cells to evade normal cellular processes that would prevent controlled cell growth and division [1].

In hematology, CML is diagnosed through a combination of physical examinations and some laboratory tests [4], such as: a) complete blood count: measures the levels of red blood cells, white blood cells, and platelets, b) blood smear: analyzes the shape, size, and number of white blood cells under a microscope [3], c) bone marrow aspiration and biopsy: a sample of bone marrow for laboratory analysis determine the presence of abnormal cells, d) cytogenetic analysis: examines the genetic material of cells to detect the Philadelphia chromosome, e) molecular testing: PCR or RT-PCR techniques are used to detect the presence of the BCR-ABL1 gene.

Hematologists face several challenges, in distinguishing CML from other types of leukemia, as the symptoms and blood cell appearance may be similar. Experts require specialized training for a thorough examination of the blood cells under a microscope to interpret the results correctly. There may be delays in diagnosis due to a lack of awareness about CML, which can lead to a delay in treatment. The lack of resources making difficult to timely diagnose CML. This can lead to the progression of the disease that make serious health outcomes. Chemotherapy and drug treatments can be harm to healthy cells and tissues, particularly chemotherapy may damage DNA in the body. An early diagnosis allows for early treatment to minimize the damage to the body. Despite recent advances, the complex and dynamic nature of CML makes it difficult to diagnose in early stages [6]. Early diagnosis is critical for preventing disease damage to the human body [10].

To the best of my knowledge, there is a lack of routine screening tests in early CML diagnosis. It is often diagnosed when symptoms start to appear or when a routine blood test shows an abnormal result. For effective diagnosing CML, molecular-diagnostic tests are important for a complete and accurate diagnosis. Molecular-level tests may be combined with traditional image-based tests. In this scenario, we proposed machine learning-based prediction system using protein amino acids sequencing information that may be benefits to individuals for early screening and monitoring. The proposed predictive system that can be used in clinical practices for effective therapeutic outcomes. Currently, we retrieved protein sequences from the Uniprot database and verify the existence of deleterious mutation in protein sequences from different data sources, including Protein Data Bank, UniProt, TCGA, and Ensemble. However, for clinical practices, protein sequencing techniques such as mass spectrometry can be utilized to identify the protein sequences in biological samples such as blood. In addition, other protein sequencing methods such as Edman degradation and N-terminal sequencing can also be used to obtain protein sequence information.

In the field of biomedical research, software tools (POLYPHEN, SIFT, and CADD) are developed that calculate the risk of mutations using protein and/or gene sequential data. The POLYPHEN tool utilizes an iterative greedy algorithm to determine the sensitivity and specificity scores of amino acid mutations in protein sequences that assess the potential impact of the mutation [1]. The SIFT tool uses protein sequence similarity and the physical properties of amino acids to predict the probability of a mutation and its potential impact. The score determines the diversity in the aligned protein sequences using the BLOSSUM

information tables [2]. The CADD tool calculates the score of single nucleotide variants 65
(SNVs) at specific genomic positions, allowing for the assessment of the potential impact 66
of the SNV [3]. These tools primarily utilize information on amino acid substitutions in 67
conjunction with the protein sequence, and the resulting scores are specific to both the 68
location and the protein. Contrary to these protein mutation risk prediction tools, we 69
have developed ML-based approach that models the discriminant/deleterious mutation 70
information in the protein sequences to predict the risk of CML. 71

The study offers a novel approach to CML identification as compared to conventional clini- 72
cal methods. The deleterious mutation analysis of frequently mutated protein sequences 73
highlights the discriminant information of amino acid sequences causing CML. The molecu- 74
lar descriptors of protein sequences are obtained using important physiochemical properties 75
of Hydrophobicity (Hb) and Hydrophilicity (Hp) of amino acid sequences. These descrip- 76
tors are expressed in various statistical and mathematical formulation-based feature spaces 77
such pseudo-amino acid-composition (PseAAC), dipeptide composition (DC), composition 78
transition distribution (CTDT), CTD composition, and amino acid composition (AAC) [11]. 79
These feature information are employed to develop several base classifiers. However, due 80
to the highly correlated nature of proteomic data, individual/base learners have limited 81
accuracy. The Multi-Layered Perceptron-based meta-ensemble optimally combines the 82
initial prediction of base learners. During model training, the proposed learning approach 83
exploited the useful discriminant information of base learners to classify CML/non-CML 84
protein sequences. Our approach has provided improved results as compared to other 85
approaches in terms of various measures. The block diagram of Meta-Ensemble classifier 86
development is given in the (Figure 1). 87

88
89

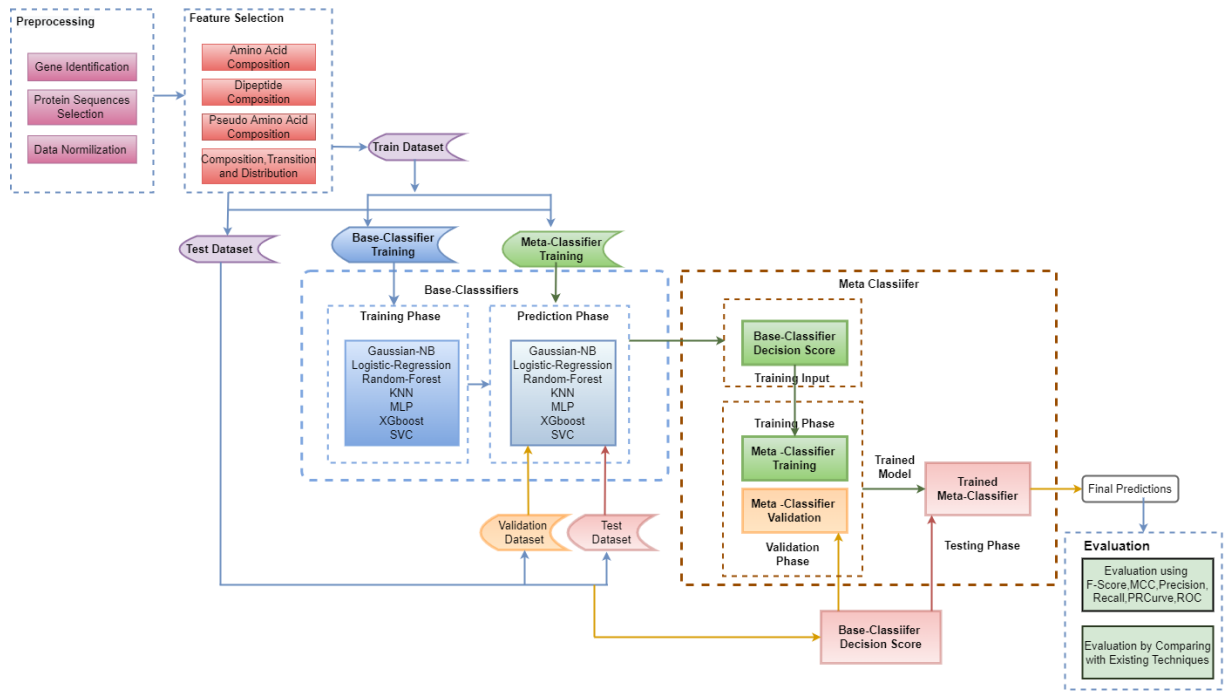

Figure 1. Block diagram of Meta-Ensemble (CML-DME) classifier development

The main contributions/innovations of this study are summarized as follows:

1. Our proposed approach has effectively utilized the discriminant information of deleterious mutations of protein sequences.
2. We have conducted a mutation analysis of protein sequences that highlighted the existence of discriminant information in protein amino acid sequences.
3. This discriminant information is transformed into molecular descriptors using the physiochemical properties of amino acids in various feature spaces.
4. The proposed approach has provided improved results in terms of various performance measures.

The remaining part of the paper is organized as follows: The second section summarizes the related work. The third section explains different parts of the proposed methodology. The fourth section focuses on the development of the meta-ensemble classifier. The fifth section presents the results and discussion. Finally, conclusions drawn from this study are given.

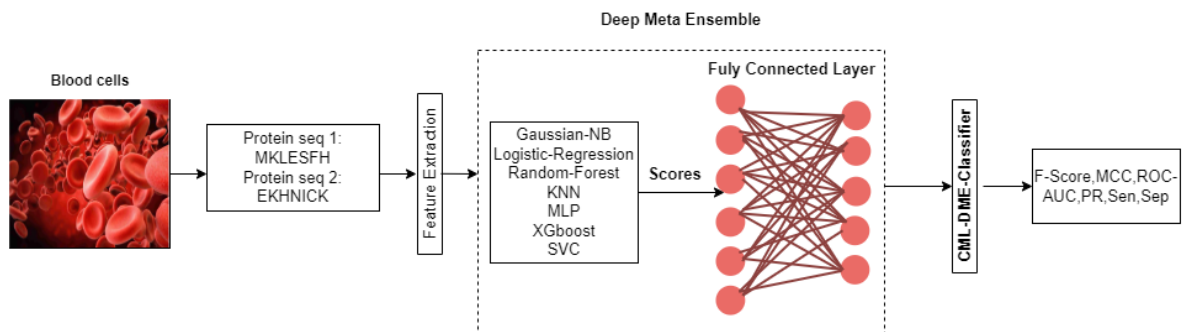

## 2. Related work

In the literature, most research on CML prediction is focused on analyzing visual images of blood cells. The image analysis techniques are used to extract features from the images and machine learning algorithms are applied to predict the presence of CML based on these features. These models are developed using histopathological images for tumor prediction, mass calculation, and segmentation. These models work in four primary steps: (i) Image Segmentation, (ii) Feature extraction, (iii) Feature selection, (iv) Classification [ 13 ].

Chan et al. have proposed a diagnostic system to segment the nucleus of white blood cells from image data. They validated their performance by using 548 nuclei retrieved from 100 images. They have investigated three feature selection algorithms by employing three classifiers and developed an automated diagnostic technique [ 14 ]. In a similar study, [ 15], classification models are developed using KNN and Naive Bayes algorithms. They reported the maximum accuracy of 92.8 % of their models for sixty sample images to classify acute lymphoblastic leukemia (ALL). In another work, [ 16 ], ensemble models are developed using three classical learning-based approaches such as SVM, RF, and ANN. They obtained

an average prediction accuracy of 75% for four types of CML. Researchers have used the 119
Weka tool to extract different features from sequential data. This tool causes ambiguity in 120
detecting proper chemical properties [17]. In another work, the author has developed an 121
XGboost classifier and reported an accuracy of 91.86 % for miRNA genes. Their results 122
demonstrated that the XGBoost model had effectively treated the imbalanced data related 123
to binary class and reported higher prediction accuracy [18]. Another research used the 124
leave-one-out cross-validation method to classify miRNA. They reported an accuracy value 125
of 83.4%. [19] 126

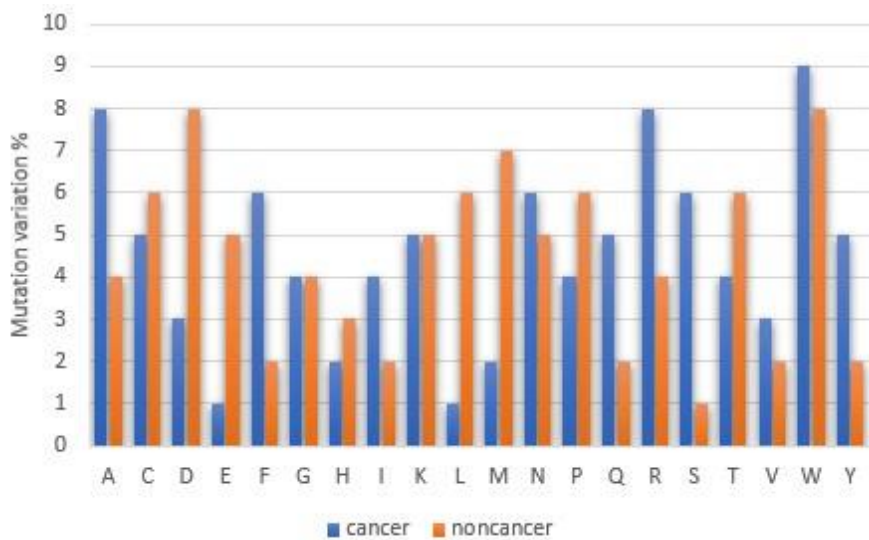

The proposed CML classifier is different from previous research. Most researchers are 127
using image analysis techniques to detect tumors in blood histopathology slides, which 128
are visual representations of blood cells. They evaluated the performance of their models. 129
Some researchers have used auto-generated features from sequential genomics data that 130
caused a high misclassification rate. In contrast, we have developed MLP-based ensemble 131

3.1. Dataset Description 157
The human protein sequences are extracted from UniProtKB [20]. This data set consists 158
of thirty thousand samples, with ten thousand sequences belonging to the positive class 159
while twenty thousand samples belong to the negative class. The online CD-HIT tool is 160
employed to eliminate homology of more than 40% within each class. This would remove 161
sequence compositional bias from impacting the machine learning model [21]. The protein 162
sequence is represented by a series of English letters representing amino acid codes. From a 163
one-dimensional perspective, a protein sequence comprises letters from the 20-letter native 164
amino acid alphabet. 165
M= A, C, D, E, F, G, H, I, K, L, M, N, P, Q, R, S, T, V, W, Y 166
167
3.2. Mutation Analysis 168
The main focus of this analysis is the identification of deleterious variations in protein
sequences. This provides evidence that there exist discriminant information exists in amino acid sequences. In the study, the protein sequences are transformed into molecular descriptors using the values of hydrophobicity and hydrophilicity of the amino acids (given in the supplementary file S1). The proposed learning approach effectively utilizes the discriminant information to classify cancer/non-cancer protein sequences. The cancer-specific variations (SNVs) in protein amino acid sequences were verified from databases such as COSMIC, TCGA, and Ensembl, while non-cancer-specific variations were validated from dbSNP. An example, a list of the most frequently deleterious mutated protein sequences is given in the supplementary data (S2).

The variation in amino acid compounds is analyzed with respect to CML/non-CML protein sequences. We divided the amino acid counts of each protein sequence into corresponding sequence lengths that help to minimize the bias created by different sequence lengths. The percentage variation of amino acids is computed as follows:

The total percent variation in amino acids between healthy and CML dataset was found to be 610.

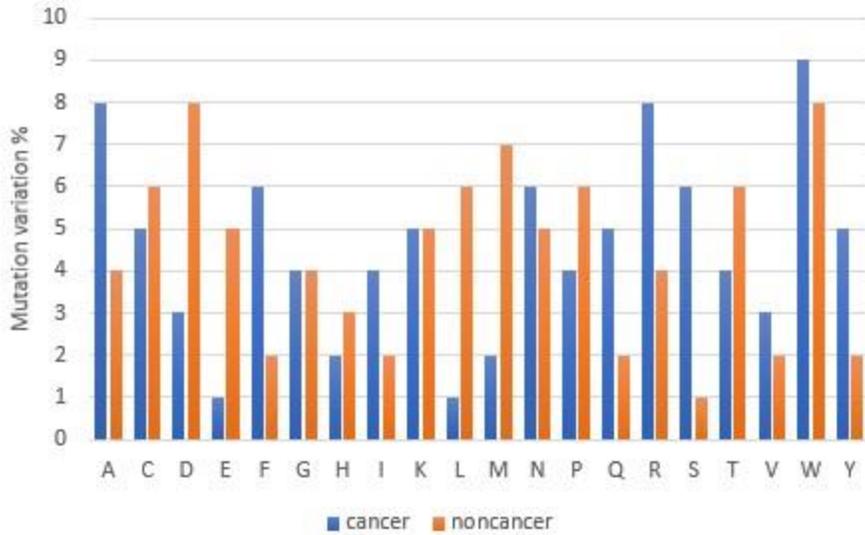

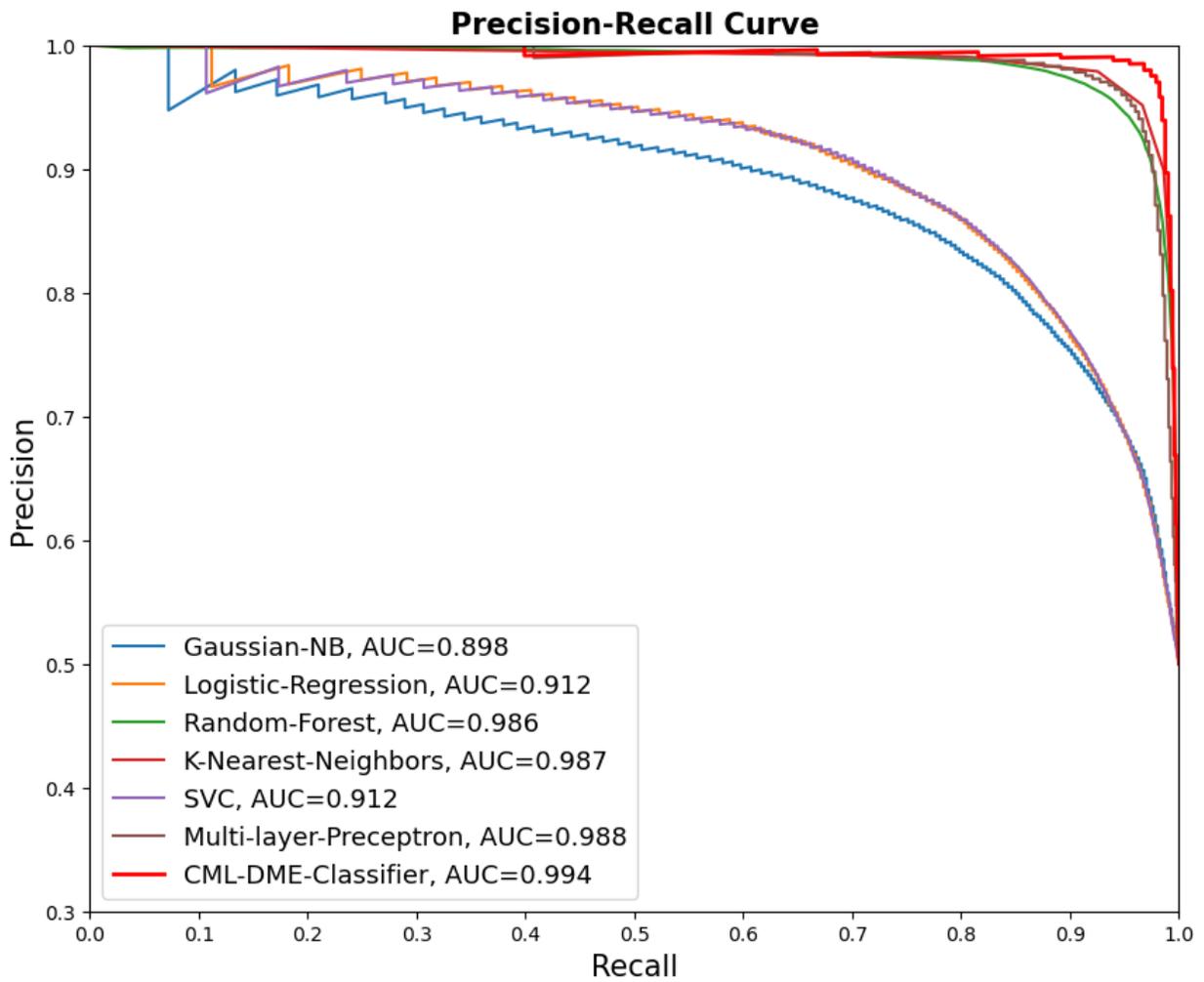

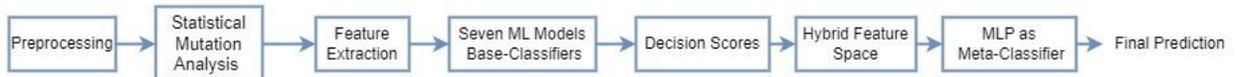

Figure 3 demonstrates the variation in amino acids of CML and healthy protein sequences. 171
Due to epigenomics and/or environmental factors, the nonsynonymous mutations in 172
amino acids raise the risk of the healthy cell to abnormal CML. In the study, our ensemble 173
approach has effectively exploited such mutations in protein sequences using the discrimi- 174
nant information in molecular descriptors expressed in various feature spaces. 175
correspondto $\alpha_i \geqslant 0$ are called support vectors. The input sequences $X_i \in \{F20, F40, F21, F21, F400\}$. 176
During SVC classifier development, We found "rbf" kernel function with optimal values of 177
gamma 2.00 and model complexity parameter (c) value 1.50. 178
179

## 4. Implementation of Meta-Ensemble Classifier 180

Since, the performance of individual base learners/classifiers does not increase beyond 181
a certain limit. Therefore, we developed a meta-ensemble classifier that optimally com- 182
bines the preliminary predicted scores of base learners. The hyper parameter regulates the 183
optimization process and the convergence of learning models. We used sklearn packages 184
to implement ML algorithms in Python for feature extraction and developing individual 185
and ensemble models. During training, the proposed ensemble model converges at 4000 186
iterations. The ensemble model is trained using 08 hidden layers with 42 random states in 187
each layer with batch size 16, relu activation function with initial learning rate 1e-5. 188
189

### 4.1. Evaluation Metrics 190

The performance of individual and meta-ensemble models is evaluated using precision 191
and recall measures. To evaluate the performance under an imbalanced dataset, we used an 192
F-score. This measure is considered an unbiased estimator that assigns equivalent weights 193
to precision and recall measures. The Matthew Correlation Coefficient (MCC) measure is 194
also used to report the performance. The MCC is another useful measure for an imbalanced 195
dataset that considers the true and false predictions for CML and non-CML classes. The 196
classifier performance is also reported at different thresholds by plotting precision-recall 197
(PR) and receiver operating characteristics (ROC) curves. 198

$$\text{Precision} = \frac{TP}{TP + FP} \quad (22)$$

$$\text{Recall} = \frac{TP}{TP + FN} \quad (23)$$

$$F - \text{Score} = \frac{2\, Recall \times Precision}{Recall + Precision} \quad (24)$$

$$MCC = \frac{(TP \times TN) - (FP \times FN)}{\sqrt{(TP + FP)(TP + FN)(TN + FP)(TN + FN)}} \quad (25)$$

$$\text{Accuracy} = \frac{(TP + TN)}{(TP + FP + FN + TN)} \quad (26)$$

199
where TP, TN, FP, and FN represent the true positive, true negative, false positive, and false 200
negative, respectively. 201

## 5. Results and Discussion 202

The individual models are developed using SVC, GaussianNB, Random Forest, LR, 203
KNN, XGBoost, and MLP algorithms. The experimental results were obtained using 10- 204

fold cross-validation technique. The experimental results of the proposed ensemble and individual ML models are presented in different feature spaces. The results show that the Dipeptide Composition (DC) feature space has improved results as compared to other feature spaces, ACC, CTDC, CTDT, and PysACC. The performance analysis is reported using performance measures of sensitivity, specificity, MCC, F-score, AUC-ROC, and AUC-PR curves. We compared the performance of the proposed ensemble with other existing ensemble models for CML classification. In the end, a temporal analysis of the proposed ensemble and base learners are given.

### 5.1. Performance Analysis of Base Classifiers

The performance of individual models is obtained in different feature spaces for CML datasets. The KNN model has obtained the overall highest Acc value of 95.9% and MCC value of 91.70% in DC feature space as compared to other feature spaces at the same time, MLP Classifier has attained the best values of AUC-ROC 98.88%, Sn 96.10%, Sp 95.6%, F-score 95.70%, and MCC 91.4% for PseAAC-S feature space, Acc value of SVM (96.99%) is better in AAC space compared to other spaces. However, it is observed that RF Classifier, in DC space has achieved the highest values of AUC-ROC 97.92%, Sn 98.26%, Sp 97.57%, F-score 97.91%, and MCC 68.01%. KNN model, in PseAAC space, has given the best AUC-ROC value of 95.45%. On the other hand, the SVC Classifier, in the PseAAC space, has provided the best AUC-ROC value of 91.20%. NB Classifier has gained a value of 98.70%, and MLP gained the highest value of 98.8% for DC feature space compared to other spaces. However, MLP has the best MCC value, 95.70%, among other feature spaces in the DC space.

### 5.2. Performance Analysis of Base Classifiers and Meta- Ensemble

Table 3 demonstrates the performance comparison of the base classifiers and the proposed ensemble in terms of F-score, precision, recall, MCC, Sn, Sp, and Acc measures, as base classifiers have reported different values of precision and recall. Therefore, we explain their F-scores in detail to report their overall performance measure. This table shows that base classifiers have obtained an F-Score in the range of (82-98)%. This highlights that base classifiers have learned the complex decision boundary of CML and non-CML protein sequences up to some limit. On the other hand, our ensemble has obtained a maximum value of F-score 98%. This highlights that, during ensemble model development, our proposed approach has given enough margin of improvement by combining the initial predictions of base classifiers. Similarly, the graphical representation of the performance of base classifiers and the proposed ensemble in terms of F-score is given in Fig. ??.

Table 2. Dimension of different well-known feature spaces.

| Serial | Feature Space | Dimension | Symbolic epresentation |
|---|---|---|---|
| 1 | ACC | 20 | F20 |

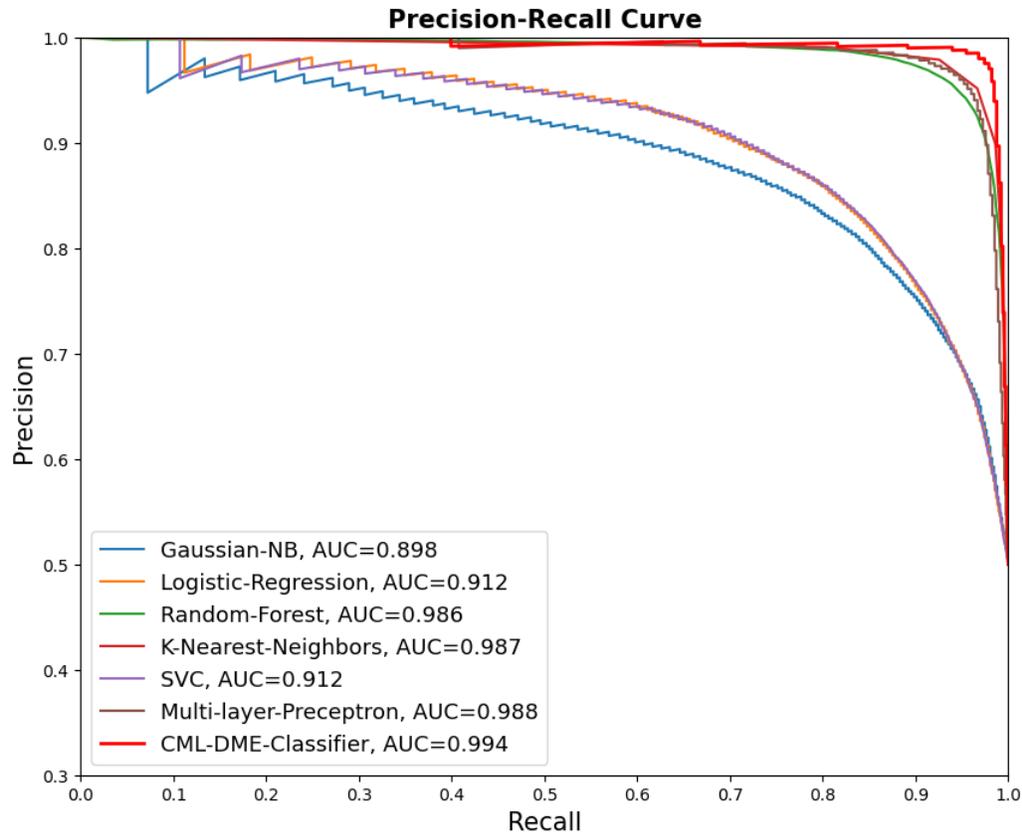

2 PseAAC 20 + λ ∗ F40
3 DC 400 F400
4 CDTD 21 F21
5 CTDC 21 F21
238

The performance comparison of the proposed ensemble with base classifiers is graph- 239
ically given in (Figure ??). This figure shows that the proposed ensemble has obtained 240
the lowest value (10) of the false-positive rate. This drastically reduced values of CML 241
as compared to base classifiers. This shows that our approach accurately predicts both 242
CML and non-CML with the lowest values of FP and FN compared to base classifiers. This 243
indicates that our ensemble has improved performance compared to base classifiers in 244
terms of FP and FN measures. 245

Table 3. Dimension of different well-known feature spaces.

Serial Feature Space Dimension Symbolic epresentation
1 ACC 20 F20
2 PseAAC 20 + λ ∗ F40

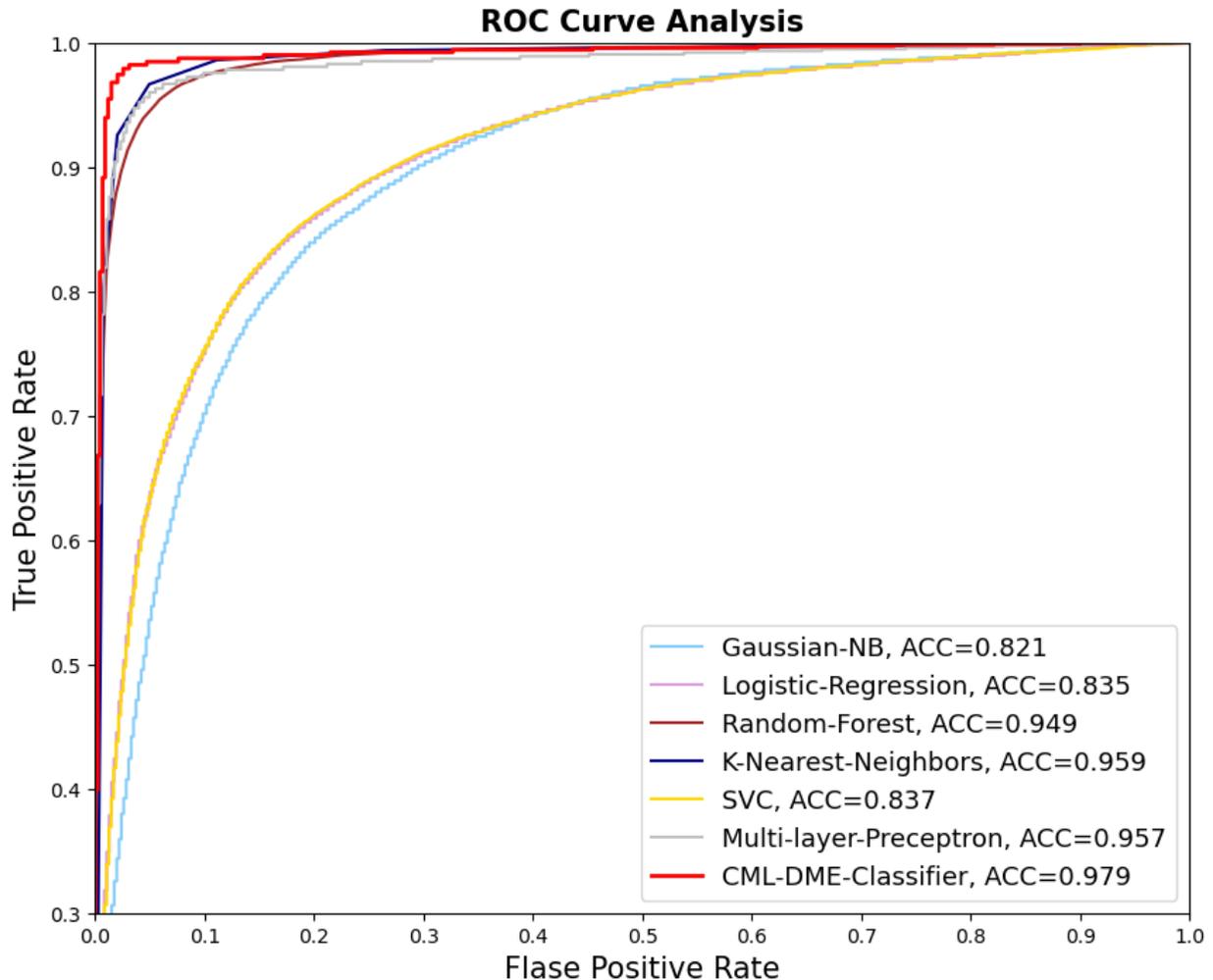

3 DC 400 F400
4 CDTD 21 F21
5 CTDC 21 F21

## 5.3. ROC Analysis 246

Further, ROC performance curves of base classifiers and the meta-ensemble are pre- 247
sented in (Figure ??). In the ROC curve, the top-left corner is the most important for 248
improving performance measures. This figure shows that, at low threshold values of 249
FPR, The TPR values of our ensemble are high as compared to base classifiers. However, 250
Gaussian-NB and logistic Regression classification have the lowest value as compared to 251
the other four Random forests, K-Nearest Neighbors, Multi-Layer Perceptron, XGBoost, 252
and SVC. This highlights the improved value of 98% of the ensemble as compared to the 253
base classifiers in terms of the AUC-ROC measure.

Table 4. Performance of meta-ensemble (CML-DME) base classifiers on test data-set
Models F-

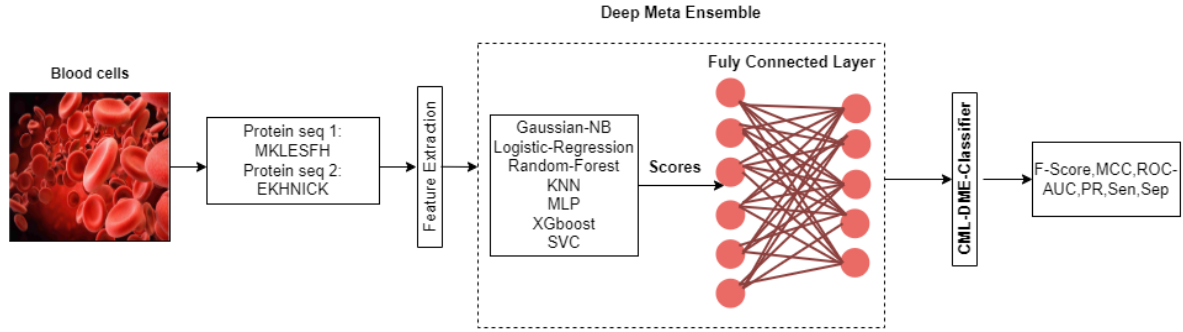

Score±S.E

PrecisionRecall MCC Sp Sn TP TN FP FN
Meta-Classifier 0.979±0.004 0.980 0.978 0.957 0.978 0.982 489 489 10 11
SVC 0.834±0.012 0.849 0.822 0.674 0.853 0.822 411 425 73 89
MLP 0.957±0.009 0.960 0.956 0.914 0.960 0.956 478 478 20 22
KNN 0.959±0.006 0.953 0.968 0.917 0.952 0.968 484 474 24 16
RF 0.948±0.008 0.952 0.946 0.896 0.952 0.946 473 474 24 27
XGboost 0.970±0.006 0.972 0.970 0.939 0.972 0.970 485 484 14 15
LR 0.834±0.012 0.842 0.834 0.671 0.843 0.830 415 420 78 85
Gaussian-NB 0.818±0.012 0.831 0.808 0.642 0.835 0.808 404 416 82 96

Note: where MCC, Sp ,Sn,TP, TN, FP, and FN represent the mathew correlation coefficient, specificity,sensitivity, true positive, true negative, false positive, and false negative, respectively.

254

### 5.4. Precision-Recall Analysis 255

Since ROC curve measurement is insufficient for an imbalanced class data set, where 256 the misclassification cost is entirely separate for positive and negative classes, the ROC 257 curve can produce unreliable results for the imbalanced class. Even if the model has very 258 low precision, it can produce a high AUC. On the other hand, the PR curve concentrates on 259 the minority group and displays the precision and detection rate (Recall). The precision of 260 the diagnosis system, in combination with its sensitivity, is critical. Precision summarizes 261 the proportion of accurately forecast positive examples. In an imbalanced dataset, a small 262 number of false positives can drastically reduce the precision of the prediction system and 263 lower the F-score. This error may have a high impact on the condition and treatment of the 264 patient. In PR analysis, the maximum AUC-PR value of 99.4% of the proposed ensemble 265 indicates that it has the highest discriminant capability to identify positive CML cases 266 from negative ones. The PR performance curves of base classifiers and the meta-ensemble 267 are presented in (Figure ??). This figure shows that, at low recall values, the precision 268 values of base classifiers are high. However, at high threshold recall values, their precision 269 performance is degraded. That is, at high recall, more data points that belong to healthy 270 cells are being classified as CML by base classifiers. On the other hand, our proposed 271 ensemble maintains its improved precision performance. According to both PR and ROC 272 curves, the proposed ensemble has good class separability across a range of thresholds, with 273 a remarkable AUC-ROC (98%) and AUC-PR (99.4%). We summarised that the proposed 274 ensemble is better as compared individual base classifiers. 275

### 5.5. Comparative analysis of proposed meta-ensemble with existing ensemble approaches 276

Researchers have developed different ensemble models employing various learning algorithms for different types of cancer classification and/or prediction systems. Each ensemble approach has its own limitation depending on the learning capability for input data. Due to the unavailability ensemble based CML prediction system, we compared the performance of proposed meta-ensemble with existing ensemble approaches for cancer classification on the different data sets. Though, this is not a fair performance comparison in terms of different measures. But this comparative analysis could be useful to highlight the importance of mutated gene in the proposed ensemble. information the valuable informative performance of different types of ensemble approaches. Table 4 demonstrates the performance of the proposed meta-ensemble with the previous ensembles. Mostly, ensemble based prediction models are developed using histopathology CML images on different feature space. In [61], cost-sensitive Naive Bayes stacking ensemble was developed using gene expression for different base learners. They have reported the ensemble maximum recall value 95.10 %. However, Our ensemble has higher value 97.80 %. Another ensemble was developed using optimal 87-dimensional feature with learning-based feature selection method for phage protein prediction. They have reported maximum accuracy of 87.23% [62]. Liang et al[18] developed a three-stage homogeneous ensemble feature selection approach for binary ovarian cancer using proteogenomics data sets. they have reported 83% prediction accuracy, with 85% sensitivity and 81% specificity. Ensemble models Extra Trees, RF, AdaBoost, and XGBoost was developed for cancer protein secretory pathway [19]. They have obtained AUR-ROC values in the range [90.0-99.0]% for RNA-seq data sets. Mostly, our our MLP-ensemble well compared to the previous ensembles in terms of accuracy, recall, and ROC measures.

The performance of the proposed ensemble is impressive for two reasons: 1) The employment of informative and discriminant features derived from the physicochemical properties of amino acids in protein sequences. These features have the discriminant capability to efficiently classify amino acids. 2) The proposed meta-ensemble with a customized MLP framework has effectively combined the predictions of individual learners at the decision level.

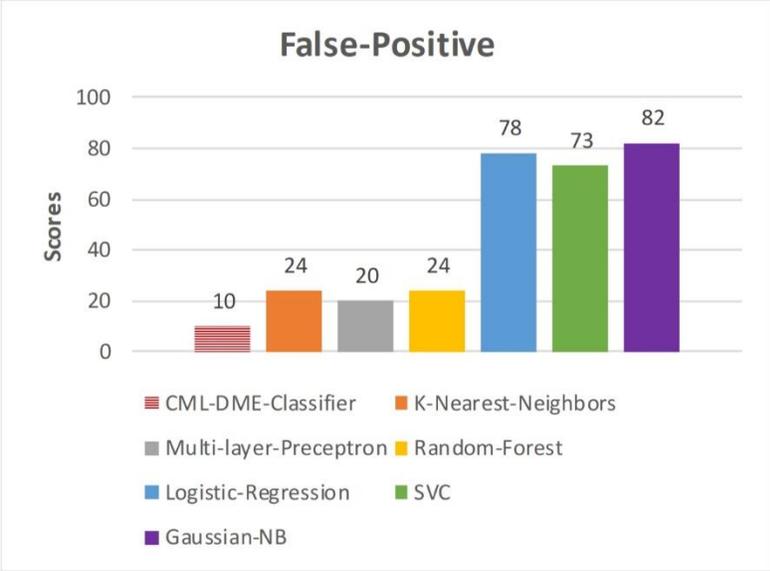

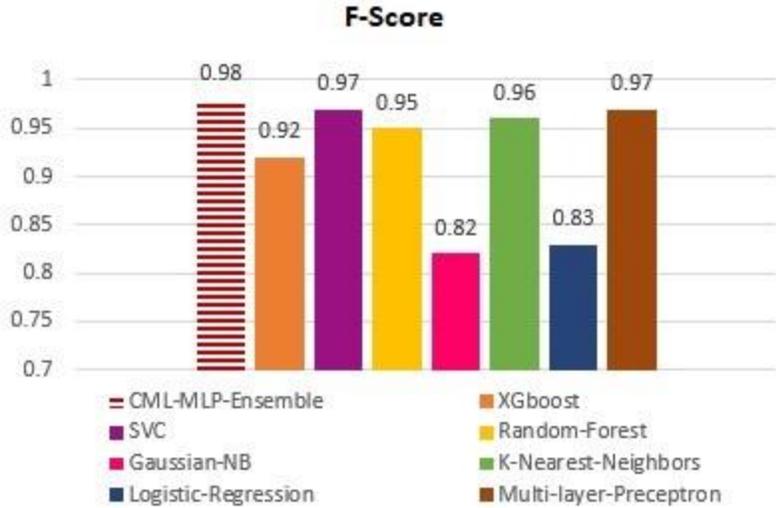

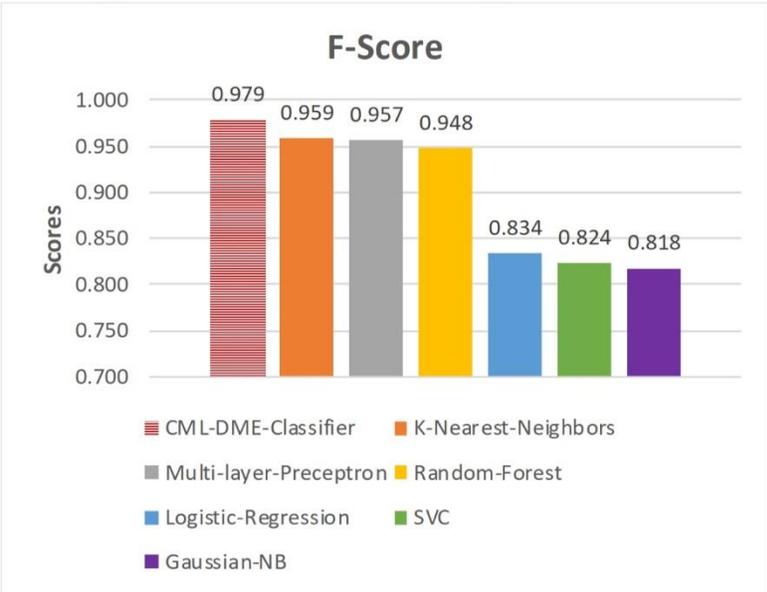

## 5.6. Temporal Cost Comparison

The temporal cost comparison of the learning algorithm is carried out in terms of computational time. The computational time of base and ensemble classifiers is computed for the testing dataset. The execution time of the learning algorithm is based on the computational environment. For ML model development and evaluation, we used a Python environment installed on 25 GB RAM with a GPU-enabled Nvidia GTX 1060 Tesla system. Table 5 presents the temporal cost comparison of base classifiers and the proposed ensemble. During the training phase, our proposed model has taken very high computational time compared to individual base classifiers. This increased computational time is due to the complex MLP neural network architecture. However, once we develop an optimal ensemble, the computational cost during the testing phase is comparable to that of conventional KNN, NB, and SVC classifiers. Our ensemble pays the temporal cost to gain other performance measures.

To summarize, the proposed ensemble approach has demonstrated improved performance compared to both individual classifiers and previous ensemble methods. The meta-level learning design is specifically designed to handle imbalanced datasets, such as the one used in this study. Although the prediction model was developed using protein data, it is anticipated that the proposed approach would perform well on genomic data after successful training.

The proposed study has some limitations as well; a.) To develop an improved prediction model, we have to select diverse types of base learners; b.) During training, we have to compute optimal values of the parameters of base learners and the hyperparameters of the proposed ensemble such as a number of hidden layers with nodes and activation functions,

## 6. Conclusions

The proposed MLP-based ensemble approach provides promising results for early risk prediction of CML using protein sequences. The deleterious mutation in the protein sequences is identified through mutation analysis that highlights the impact of deleterious mutations on protein structure and function. This approach offers a potential biomarker by exploiting the deleterious mutations of protein sequences. The performance analysis showed that the meta-ensemble outperforms base classifiers in terms of accuracy, recall, and ROC measures, with an improved AUC-ROC value of 98%. The performance analysis, in different feature spaces, showed that the DC feature space is more effective. The F-score of base classifiers obtained is in the range of (82-98)%. The PR analysis revealed the highest AUC-PR value of 99.4% for the meta-ensemble, indicating its robustness and generalization for the imbalanced CML dataset. From the ROC analysis, the meta-ensemble outperforms base learners in terms of accuracy, recall, and ROC measures. With the proposed ensemble, an improved AUC-ROC value of 98% as compared to the base classifiers was obtained. The approach effectively exploits the discriminant information of mutated proteins in the proteomic domain and has a high discriminant capability for CML case identification.

In future research, we plan to further enhance the proposed framework by incorporating the nonsynonymous gene mutation profile and analyzing its significance biological. This is because nonsynonymous gene mutations have a significant impact on protein structure and function, and the analysis of these mutations can provide valuable insight into the biological processes underlying CML.

Author Contributions: MH, MB, TM, AM and AK conceptualized the project. MH and AM curated 352
the dataset. MH, MB and TM analysed the data. MH and TM generated the feature space. MH 353
implemented the classifiers. MH and MB trained, tested and validated the classifiers. MH, MB and 354
TM analysed the results. MH and TM visualized the performance of the classifiers. MH, MB, TM, 355
AM and AK reviewed and edited the manuscript. AM and AK supervised the research. All authors 356
have read and agreed to the published version of the manuscript. 357
Data Availability Statement: A supplementary file is provided with some of the data. The com- 358
plete dataset generated during and/or analyzed during the current study are available from the 359
corresponding author on reasonable request. 360
Funding: The research was not conducted as part of any project, nor did the authors receive any 361
funding. 362
Conflicts of Interest: The authors declare no conflict of interest. 363